# Extended Equivalence Principle: Implications for Gravity, Geometry and Thermodynamics[*]

*C Sivaram*

Indian Institute of Astrophysics, Bangalore, 560 034, India

Telephone: +91-80-2553 0672; Fax: +91-80-2553 4043

e-mail: sivaram@iiap.res.in

*Kenath Arun*

Christ Junior College, Bangalore, 560 029, India

Telephone: +91-80-4012 9292; Fax: +91-80- 4012 9222

e-mail: kenath.arun@cjc.christcollege.edu

**Summary:** The equivalence principle was formulated by Einstein in an attempt to extend the concept of inertial frames to accelerated frames, thereby bringing in gravity. In recent decades, it has been realised that gravity is linked not only with geometry of space-time but also with thermodynamics especially in connection with black hole horizons, vacuum fluctuations, dark energy, etc. In this work we look at how the equivalence principle manifests itself in these different situations where we have strong gravitational fields. In recent years the generalised uncertainty principle has been invoked to connect gravity and curvature with quantum physics and now we may also need an extended equivalence principle to connect quantum theory with gravity.

---

[*] This paper received the Honourable Mention in 2012 Awards for Essays on Gravitation

The equivalence principle was formulated by Einstein in 1907 in a bid to extend the notion of inertial frames (with their uniform motion) to accelerated frames, thereby bringing in gravity [1]. Einstein's remarkable insight in postulating that gravity is locally equivalent to an accelerated frame and can be transformed away enabled the understanding of the then empirical observation that all bodies (of whatever constitution or mass) fall at the same rate in a given gravitational field, such as that of the earth.

The absence of forces in a freely falling system was a consequence and its practical realisation of 'weightless' astronauts was universal a few decades later in innumerable spaceflights!

However the connection of gravity with geometry, i.e. interpretation entirely in terms of the parameters of space-time geometry was formulated only a few years later when Einstein made several attempts to setup the complex field equations of general relativity with the mathematician Grossmann (1912-16) [2, 3]. Thus the equivalence principle as such just gave a clue in understanding the 'equality' between inertial and gravitational mass and the empirical universality of free fall. This was one reason why relativists like J. L. Synge remarked that the equivalence principle really was only the 'midwife' in initiating this first step thereafter not playing much role in the complex Riemann geometric formulation and other aspects. He even felt the 'midwife' be 'buried with due honours'! [4]

Again in recent decades, it has been realised that gravity is linked not only with geometry but also with thermodynamics [5, 6]. In cosmology there are phenomena like dark energy, which involve quantum vacuum fluctuations leading to effective cosmological constant etc. How does the equivalence principle manifest itself in these situations? [7, 8]

Moreover we know that the equivalence principle is strictly 'local', only we can transform away the gravitational field locally. A sufficiently large laboratory or 'framework' can detect tidal forces which measure the gradients in the field which are the truly 'measureable' quantities.

In the geometry picture, an infinitesimal region of a curved space can be treated as Euclidean or flat and free of gravitational fields. By introducing flat space co-ordinates 'locally' we can have a gravity-free inertial frame, where the laws of special relativity are valid. Once can also have the strong equivalence principle where the laws of physics are valid in all locally freely falling frames.

*How local is local?* In a strong gravitational field in the vicinity of a compact object even infinitesimally separated regions can feel large tidal forces. A one metre object can be torn up near a neutron star, while in the vicinity of a primordial black hole; say at a distance of a fermi ($\sim 10^{-15}m$) from a $10^{12}kg$ primordial black hole, even at an atomic nucleus can be torn apart; the tidal gravitational force being greater than the nuclear force (similar situations would have been present in the very early universe):

$$\frac{GM_{BH}m_p r}{d^3} \gg \text{Nuclear binding force} \qquad \ldots (1)$$

(for $M_{BH} = 10^{12} kg$, $m_p = 10^{-27} kg$, $r \approx d = 1f$ )

So in such strong gravitational fields, even subatomic regions cannot be free of gravity. This is why, strictly speaking Lorentz invariance (flat space and gravity free symmetries) hold only for point particles and point-particle interactions. To detect gravitational waves (i.e. curvature variations) one has to study time varying separation between pairs of particles, i.e. need extended body to measure gravity.

In a way one can naively understand why string theory accommodates gravity, because an extended object (however infinitesimal) can register a 'tidal' (true gravitational force!) especially in strong gravitational fields (such as those at the Planck scale!).

Similar arguments hold for curvature. In a strongly curved surface, we can consider only a really infinitesimal region as 'flat', as can be seen by the triangular deviation: [9]

$$\Delta\theta = \frac{A}{R^2} \qquad \ldots (2)$$

(*A* is the area of the infinitesimal region on surface and *R* is the radius of curvature.)

Thus in strongly curved space (small *R*), there would be large deviations from flat space ($\Delta\theta$ large) even in infinitesimally small regions which is just the *equivalent* way of saying that in a strong gravitational field, there would be large tidal forces even over infinitesimal distances. (For primordial black holes or super-strings the equivalence principle in the context of such strong gravitational fields (or large curvature) in their vicinity ceases to have much operational meaning!)

Thermodynamics has now entered into discussion of geometry and gravitational fields. A black hole horizon, with a constant surface gravity, has an equivalent black body temperature proportional to the surface gravity (inversely as the horizon radius). So also an accelerated observer finds himself immersed in a thermal bath of temperature proportional to the acceleration (equivalently surface gravity!). By the same equivalence of gravity and geometry, a curved space (with a horizon) is **hot** and is associated with a temperature proportional to the curvature!

Thus we need to have a generalised equivalence principle connecting curvature, acceleration, gravity and thermodynamics. Curvature and thermodynamics were fully absent in the initial formulation of the equivalence principle by Einstein. An inertial frame cannot be associated with a temperature, curvature or gravity!

High accelerations, strong gravitational fields and strongly curved spaces are all associated with high temperatures and are all equivalent! [10]

An observer in a uniform accelerated frame, a black hole horizon with constant gravitational field and a space with constant curvature are all associated with a uniform temperature *T*. [11] (equivalence universality of these systems!)

$$T = \frac{\hbar a}{k_B c} = \frac{\hbar c \sqrt{\Lambda}}{k_B} \qquad \ldots (3)$$

(*a, g, Λ* are acceleration, surface gravity and curvature)

The extended equivalence principle would associate a temperature (and other thermodynamic parameters) with accelerated frames, curved spaces and gravitational fields of exotic objects such as black holes, strings, etc. An inertial frame cannot radiate (so no forces, continues with uniform velocity or remains at rest, whatever be its mass).

A non-zero *T*, implies an acceleration, same for all particles and an entropic force proportional to mass [10]. An accelerated frame would lose mass, by emitting radiation, which in turn would make it accelerate more and raise the temperature. This is exactly what happens to a black hole horizon emitting Hawking radiation. It progressively decreases in mass, increasing in temperature, radiation at even higher rates (i.e. a system with negative specific heat). Associating the horizon of an accelerated observer by the curvature radius $\sim c^2/a$, all the formulae are equivalent, if *a* is the equivalent surface gravity (of the corresponding black hole).

Entropy is again associated with curvature [12]:

$$\text{Entropy} \times \text{curvature} = \text{Constant} \quad \ldots (4)$$

(For an accelerated observer, this is proportional to $\sim a^2/c^4$)

Entropy (or information lost) would be the total number of quanta emitted in the complete decay of both the black hole horizon or the accelerated observer's horizon (for black holes this is $\sim GM^2/\hbar c$)

The minimum entropy $\sim k_B$, has a maximum curvature $\sim c^3/\hbar G$. [12]

To reduce the temperature of a black hole, we have to keep adding matter (to make it grow!) to it, as $T \propto \dfrac{1}{M}$.

As $M \to \infty$, for an indefinitely large mass, $T \to 0$. This can be regarded as a manifestation of the third law of thermodynamics, i.e. you need infinite number of steps to reach absolute zero (i.e. keep adding particles to the black hole to asymptotically reach $T = 0$).

Again for $T = 0$, the horizon radius $R_H \to \infty$, so that one cannot increase entropy once $T = 0$, i.e. entropy changes vanish as $T \to 0$.

For a black hole with indefinitely large mass, $M \to \infty$, the curvature tends to vanish $(\kappa \to 0)$, surface gravity vanishes, so it is equivalent in all senses to a ***flat space-time***! So the horizon geometry of an indefinitely massive black hole is indistinguishable from flat space!

Although as pointed out in references [12, 13], the interior phase space of any black hole (of whatever mass) is just a quantum of phase space $\hbar^3$ containing just one photon (i.e. lowest interior entropy). Again the equivalence between temperature and curvature (and gravity) suggest that 'curvature flows' (in the sense of Ricci flow) can be written like a thermal (or diffusion) gradient equation for heat flow: ($\kappa$ is the equivalent conductivity) [10]

$$\frac{\partial R}{\partial t} = \kappa \nabla^2 R \qquad \ldots (5)$$

Now for a uniform temperature (at surface) for a constant 'conductivity', the solution of the conduction equation (for temperature) is:

$$T = \kappa \left( r_0^2 - r^2 \right) \qquad \ldots (6)$$

which is $\propto \kappa r_0^2$, for $r_0 >> r$.

Now for a space of constant curvature, we also have an energy proportional to $r^2$, i.e. the de-Sitter solution where the potential grows as $r^2$. There are several such interesting analogies. This perhaps also modifies the propagation of gravitational or 'curvature' waves over large distances. This is being currently investigated.

Also the string-black hole equivalence [14] implies that the horizon of a black hole is equivalent to a super-string with a tension identically $c^2/G$. String thermodynamics also follows from this equivalence. The horizon surface can be generated by a string of the above tension.

Again in Einstein's field equations, the gravitational constant does not manifest in the 'vacuum' field equations, i.e. the empty space field equations. Thus $G_{\mu\nu} = 0$, does not tell anything about the gravitational coupling constant *G*. This is equivalent to an inertial frame where $F = 0$. Only in the non-inertial case where we have $F = ma$, does the mass manifest itself as a measureable parameter.

Similarly in the presence of matter (energy-momentum) the field equations invoke the gravitational constant $\kappa$, through $G_{\mu\nu} = \kappa T_{\mu\nu}$. The energy-momentum can be generated by all fields (including gravitational fields). Just as all forms of mass are equivalent, all fields causing gravity (to curve space) have the same universally equal gravitational coupling constant, $\kappa = 8\pi G$!

In the sense of Sakharov, $\kappa$ is the metrical elasticity of space-time, so all interactions causing gravity by their contributions to the energy-momentum tensor (including quantum vacuum fluctuations) curve space with the same elasticity coefficient perhaps suggesting a microscopic origin for *G*.

This profound equivalence is still empirical and could be better explained by a unified theory of space-time with all basic interactions! [15]

We finally remark that the generalised uncertainty principle [16, 17] has been invoked in recent years to connect gravity and curvature with quantum physics and now we also need a generalised or extended equivalence principle to connect quantum theory with gravity.

**References:**

1. Einstein A (1907) Jahrbuch der Radioaktivität und Elektronik 4, 252
2. Einstein A and Grossmann M (1914) Zeitschrift für Mathematik und Physik, 62, 225
3. Einstein A and Grossmann M (1914) Zeitschrift für Mathematik und Physik, 63, 215
4. Synge J L, Talking about relativity, North-Holland Pub. Co.: Amsterdam (1970)
5. Susskind L (1995) J. Math. Phys. 36, 6377
6. t Hooft G': in A. Aly, J. Ellis et al. eds., World Scientific, (1993)
7. Sivaram C (2001) Gen. Rel. Grav., 33, 1
8. Sivaram C (2007) Chapter 29 in Foundations of Sciences (PHISIC-CSC, New Delhi) Aspects of Dark Energy and Black Hole Entropy, pp.550-65
9. Misner C, Thorne K and Wheeler J, Gravitation, Freeman and Co.: San Francisco (1973)
10. Sivaram C (2010) Gravity of ITS from BITS VIA Information, Holography and Vacuum Energy, Received honourable mention at 2010 Essays in Gravity

11. Sivaram C (1995) in "Currents in High-Energy Astrophysics", NATO-ASI Series, eds. Shapiro M et al, Kluwer Academy Press: Dordrecht, p.177; Sivaram C (1994) Astrophys. Space Sci., 219, 135

12. Sivaram C and Arun K (2009) Int. J. Mod. Phys. D, 18, 2167

13. Sivaram C (2004) Asian J. Phys., 13, 293

14. De Sabbata V and Sivaram C, Spin and Torsion in Gravitation, World Scientific (1994)

15. Sivaram C (1994) Int. J. Theor. Phys., 33, 2407

16. Sivaram C, Arun K and Samartha C A (2008) Mod. Phys. Lett. A, 23, 1470

17. Sivaram C (1990) Astrophys. Space Sci., 167, 335